\documentclass[fleqn,usenatbib]{mnras}

\usepackage{newtxtext,newtxmath}
\usepackage[T1]{fontenc}

\DeclareRobustCommand{\VAN}[3]{#2}
\let\VANthebibliography\thebibliography
\def\thebibliography{\DeclareRobustCommand{\VAN}[3]{##3}\VANthebibliography}

\usepackage{graphicx}	% Including figure files
\graphicspath{{figures/}}
\usepackage{amsmath}	% Advanced maths commands

\usepackage{multirow}
\usepackage{longtable}

\usepackage[capitalise]{cleveref}
\crefname{figure}{Fig.}{Figs.}
\Crefname{figure}{Fig.}{Figs.}

\usepackage{acro}
\DeclareAcronym{GR}{
	short = GR,
	long  = general relativity
	}
	
\DeclareAcronym{BH}{
	short = BH ,
	long  = black hole
}

\DeclareAcronym{BBH}{
	short = BBH ,
	long  = binary black hole
}

\DeclareAcronym{NSBH}{
	short = NSBH ,
	long  = neutron star black hole
}

\DeclareAcronym{GW}{
	short = GW ,
	long  = gravitational wave
}

\DeclareAcronym{CBC}{
	short = CBC,
	long  = compact binary coalescence
}

\DeclareAcronym{SNR}{
	short = SNR,
	long  = signal-to-noise ratio
}

\DeclareAcronym{PSD}{
	short = PSD,
	long  = power spectrum density
}

\DeclareAcronym{FAR}{
	short = FAR,
	long  = false alarm rate
}

\DeclareAcronym{HSLM}{
	short = HSLM,
	long  = highly spinning light mass
}

\def\msun{$M_\odot$}

\title[Search for High Spin Light Compact Binaries]{Targeted Search for Gravitational Waves from Highly Spinning Light Compact Binaries}

\author[Y.-F. Wang and A.H. Nitz]{
Yi-Fan Wang,$^{1,2}$\thanks{E-mail: yifan.wang@aei.mpg.de}
Alexander H. Nitz$^{3}$
\\
$^{1}$Max-Planck-Institut f{\"u}r Gravitationsphysik (Albert-Einstein-Institut), D-30167 Hannover, Germany\\
$^{2}$Leibniz Universit{\"a}t Hannover, D-30167 Hannover, Germany\\
$^{3}$Department of Physics, Syracuse University, Syracuse, NY 13244, USA
}

\date{Accepted XXX. Received YYY; in original form ZZZ}
\pubyear{2023}

\begin{document}
\label{firstpage}
\pagerange{\pageref{firstpage}--\pageref{lastpage}}
\maketitle

\begin{abstract}
Searches for gravitational waves from compact-binary mergers, which to date have reported $\sim100$ observations, have previously ignored binaries whose components are both consistent with the mass of neutron stars $(1 M_\odot$ to $ 2 M_\odot)$ and have high dimensionless spin $>0.05$. While previous searches targeted sources that are representative of observed neutron star binaries in the galaxy, it is already known that neutron stars can regularly be spun up to a dimensionless spin of $\sim0.4$, and in principle reach up to $\sim0.7$ before breakup would occur. Furthermore, there may be primordial black hole binaries or exotic formation mechanisms to produce light black holes. In these cases, it is possible for the binary constituent to be spun up beyond that achievable by a neutron star. A single detection of this type of source would reveal a novel formation channel for compact-binaries. To determine if there is evidence for any such sources, we use PyCBC to conduct a targeted search of LIGO and Virgo data for light compact objects with high spin. Our analysis detects previously known observations GW170817 and GW200115; however, we report no additional mergers. The most significant candidate, not previously known, is consistent with the noise distribution, and so we constrain the merger rate of spinning light binaries.
\end{abstract}

\begin{keywords}
gravitational wave -- compact binary coalescence -- neutron star
\end{keywords}

\section{Introduction} \label{sec:intro}

The detection of gravitational waves offers a distinct avenue for investigating compact binary systems consisting of black holes or neutron stars, which are complementary to means of electromagnetic telescopes. 
Up to the completion of the third observation run in 2020, ninety gravitational wave events have been reported by the LIGO and Virgo Collaboration in the Gravitational Wave Transient Catalog-3 \citep[GWTC-3,][]{2111.03606}.
Additional events are reported by 4-OGC \citep{10.3847/1538-4357/aca591} and the IAS catalog \citep[e.g.][]{2201.02252}.
These detections have provided significant contributions to our understanding of the black hole and neutron star population, and the experimental verification of general relativity, to name a few \citep[e.g.][]{2111.03634,2112.06861}.
The fourth observational run began in May of 2023 with enhanced sensitivity; Advanced LIGO \citep{10.1088/0264-9381/32/7/074001} and Advanced Virgo \citep{10.1088/0264-9381/32/2/024001} target a horizon distance 170 Mpc and 120 Mpc \citep{1304.0670}, respectively. 
Additionally, a fourth detector, KAGRA \citep{10.3390/galaxies10030063}, with a current target horizon distance of 5 Mpc, has been incorporated into the joint observation.

The most sensitive gravitational-wave searches for compact binary coalescence employ matched-filtering by correlating the data stream with a pre-established template bank \citep{10.1103/PhysRevD.46.5236, 10.1103/PhysRevD.49.1707}.
The efficacy of a matched-filtering search depends crucially on the parameter space of a bank. 
Outside the bank, the level of sensitivity would decrease, albeit with the possibility of some compensation due to waveform parameter degeneracy.
The prior searches, which reported nearly 100 observations, target a range of binary component masses from one to a few hundreds solar masses.
The spin of compact objects can be characterized by the dimensionless parameter $\vec\chi_{1/2} = \vec s_{1/2} / m_{1/2}^2$, where $\vec s_{1/2}$ and $m_{1/2}$ are the angular momentum and mass of a binary component ($G=c=1$).
For objects with mass consistent with a neutron star, specifically $1-2~M_\odot$, the amplitude of dimensionless spin was assumed to be from 0 to 0.05 in the direction parallel or antiparallel to the orbital angular momentum.
The spin constraint is relaxed to $\sim$ 0.9 for component masses greater than 2 $M_\odot$, under the assumption that they may be black holes.

 \begin{figure*}
%\begin{center}
\includegraphics[width=16cm,height=4cm]{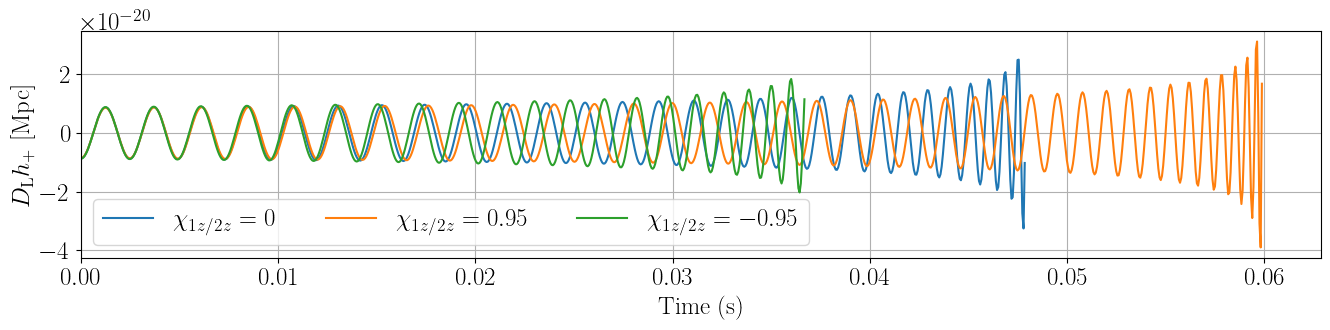}
\caption{Illustrations of dynamical imprints from varying effective spins, which are chosen to be 0 and $\pm$ 0.95 using the waveform model \textit{TaylorT4}.
The vertical axis denotes the multiplication of luminosity distance $D_L$ and the plus polarization $h_+$ in the unit of Mpc. 
The component mass in detector frame is 1.4-1.4 \msun{}.
For better visualization, we only evolve the waveform from 400 Hz, which contains the last several cycles before the inner stable circular orbit of the binary.}
\label{fig:waveform}
%\end{center}
\end{figure*}

The choice of spin constraint \citep{nitz2015effect} was based on the observation that the spins of all known galactic double neutron stars are consistently less than 0.05~\citep{10.1103/PhysRevD.98.043002}.
PSR J1946+2052 is the fastest spinning neutron star in a double neutron star binary with a period of 17 ms \citep{10.3847/2041-8213/aaad06}. 
If it had a mass of 1.4 $M_\odot$ and a radius of 11 km, the magnitude of its dimensionless spin, $|\vec\chi|$, would be approximately $0.03$.
Consequently, these types of neutron stars would be detected by previous gravitational-wave searches when they merge.
However, one of the fastest spinning neutron stars observed to date,  PSR J1748-2446ad, has a frequency of 716 Hz in~\citep{2006Sci...311.1901H}, corresponding to $|\vec\chi|\sim0.4$. 
Neutron stars could sustain spin up to a maximum of $\sim$ 0.7, beyond which they would be torn apart, with the details depending on the equation of state~\citep{10.1088/0004-637X/728/1/12}. 
If it is possible to assemble highly spinning neutron stars and have them merge within the LIGO observing window, these may be missed by current analysis. 
Observing such a source would inform us about the available formation channels~\citep{10.1103/PhysRevD.98.043002} and that they must include an efficient mechanism for the accumulation of angular momentum, most likely through accretion~\citep{10.1086/175268}. 

The potential also exists that there are light compact binaries that may not be neutron stars, thus the neutron star spin upper limit would not apply.
GW170817 remains the only observation with corroborated electromagnetic emission~\citep{1710.05833}. 
Due to the observed gamma-ray burst~\citep{1710.05834} and kilonova~\citep{1710.05836} it is clear that at least one binary component and most likely both are neutron stars. 
GW190425 was observed with chirp mass $1.44^{+0.02}_{-0.02}~M_\odot$  \citep{2001.01761}, which lies within the neutron star merger mass range. 
The component spin, is bound to $<0.33$ with a prior of upper limit 0.89 (note this high limit prior is only used in the parameter estimation not during the search). 
The association as a binary neutron star is based solely on its chirp mass as there was no confirmed electromagnetic counterpart and finite-size effects can not be sufficiently constrained~\citep{2001.01761} to exclude that the binary is composed of black holes.
Several merger observations have components which span the ``lower mass gap'' between 2-5 solar masses~\citep[e.g.][]{2006.12611}. 
The possibility of the black hole mass distribution extending even lower can be confirmed by the existence of a highly spinning binary.
Additionally, highly spinning compact objects with mass $[1,2]~M_\odot$ may indicate the presence of primordial black holes \citep[e.g.][]{2007.06481}, which were formed by the direct collapse of overdensity in the early universe. 
In the context of spherical collapse, the primordial black holes are not expected to possess spin at birth.
However, this is not the case for asymmetric collapse \citep{1903.01179}.
Primordial black holes may also accumulate spin through hierarchical mergers~\citep{10.1103/PhysRevD.107.063035}.
By detecting gravitational waves from these objects, we can probe the conditions of the early universe and gain a better understanding of how these objects were formed.
Other exotic formation mechanisms for light compact binaries such as quark stars may also exist \citep{10.1088/0004-637X/728/1/12} and transcend the imposed low spin limit.

In this study, driven by the fact that previous searches limited the component spin to less than 0.05 for low-mass binaries, we perform a targeted search for gravitational waves with a broader spin range up to 0.95 in this mass range.
We use the open-sourced \texttt{PyCBC} \citep{pycbc-github} toolkit to analyze the entire public LIGO and Virgo data from 2015 through 2020 \citep{1912.11716,2302.03676}.
Our analysis identified the already known binary neutron star observation GW170817 and the neutron star black hole event GW200115 \citep{2106.15163}. 
However, no additional significant gravitational wave events were detected in the data.
Nevertheless, this result suggests that the presence of such systems is rare, allowing us to place the first upper limits on the rate of highly spinning compact binaries with mass consistent with neutron stars in the local universe.

\section{Template bank mismatch against high spin light compact binaries}\label{sec:mismatch}

The effect of spin on the dynamics of compact binary inspirals is predominately encoded in the effective spin parameter \citep{10.1103/PhysRevD.47.R4183,10.1103/PhysRevLett.106.241101}, which is defined as 
\begin{equation}
\chi_\mathrm{eff} = \frac{m_1\chi_{1z}+m_2 \chi_{2z}}{m_1+m_2},
\end{equation}
where $\chi_{1/2z}$ is the component spin in the orbital angular momentum direction.
This work only considers the case where the component spin is aligned (anti-aligned) with the orbital angular momentum.
One notable characteristic is that a positive effective spin causes a binary to undergo a longer inspiral, and a negative, antialigned, effective spin has the opposite effect. 
We visualize this effect by plotting three waveforms in \cref{fig:waveform} using the post-Newtonian waveform approximant \textit{TaylorT4}, which is a time domain model derived from Taylor expansion up to 3.5 post-Newtonian order \citep{0710.0158, 0907.0700}.
To illustrate the waveform from the last several cycles before the merger, we start to evolve the waveform from a frequency of 400 Hz.
The waveform terminates at the last inner stable circular orbit. 
It is clearly shown a substantial dephasing is caused by high effective spin compared with the zero-spin case. Hence, the matched-filtering \ac{SNR} will be severely reduced if the correct spin parameter is not considered.

We quantify the degree of \ac{SNR} loss by comparing the template bank used by a previous search against simulated signals with varying chirp mass and effective spin.
Typically the following inner product is used to measure the separation between two gravitational wave templates
\begin{equation}
(h_1,h_2) = 4\Re \int \frac{h_1(f) h^\ast_2(f)}{S_n(f)} df
\end{equation}
where $h_1$ and $h_2$ are signals or gravitational wave templates in the frequency domain, and $S_n(f)$ is the one-sided noise power spectral density.
The overlap is characterized by taking the normalization into account
\begin{equation}
\mathcal{O}(h_1,h_2) = \frac{(h_1,h_2)}{\sqrt{(h_1,h_1)(h_2,h_2)}}.
\end{equation}
Two templates may be different up to an adjustable coalescence time and phase in the frequency domain. Thus the match function between two waveforms is given by maximizing over the offset of coalescence time $t_c$ and an overall phase $\phi_c$
\begin{equation}\label{eq:match}
\mathcal{M}(h_1,h_2) = \max_{t_c,\phi_c}\left(\mathcal{O}(h_1,h_2e^{i(2\pi f t_c -\phi_c)})\right).
\end{equation}

\begin{figure}\includegraphics[width=\columnwidth]{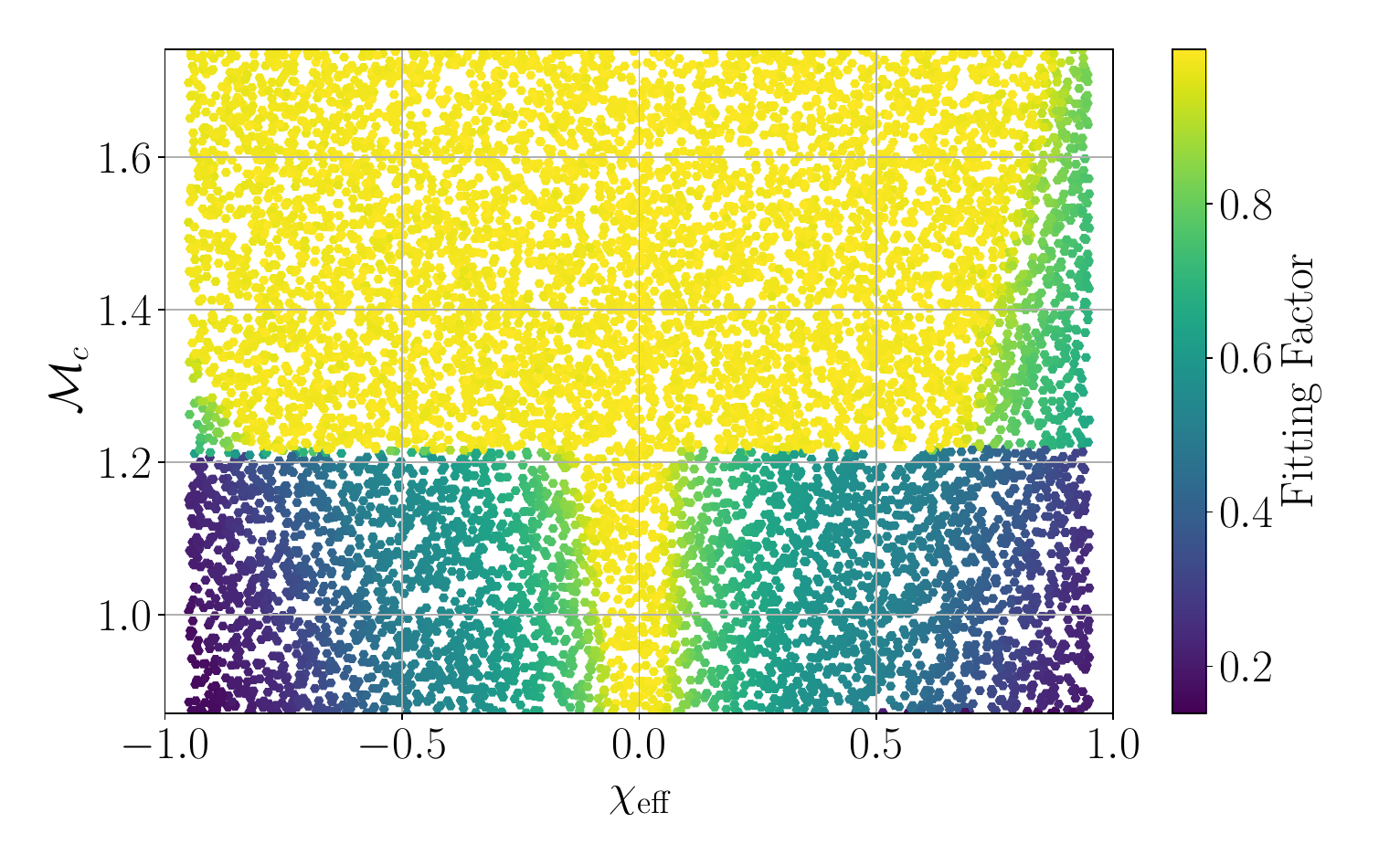}
\caption{Results of fitting factor from a comparison of simulated waveforms and the template bank from 4-OGC \citep{10.3847/1538-4357/aca591} as a representative example of prior searches.
$\mathcal{M}_c \sim 1.2$ \msun{} clearly marks a boundary between the results because this corresponds to the border of the parameter space between binary neutron star and neutron star black hole, where the former is allowed to possess higher effective spin.}
\label{fig:ff}
\end{figure}

We compare the template bank used in 4-OGC \citep{10.3847/1538-4357/aca591} against simulated signals with high spin.
We generate multiple simulated waveforms with equal component mass but varying chirp mass $\mathcal{M}_c \equiv (m_1m_2)^{3/5}/ (m_1+m_2)^{1/5}$ distributed from 1 to 2 $M_\odot$ and equal spin but varying effective spin $\chi_\mathrm{eff}$ from -0.95 to 0.95.
\cref{fig:ff} illustrates the fitting factor, defined as the match between a simulated signal and the best-fitting template in the bank. 

There is a clear separation for $\mathcal{M}_c \sim 1.2 M_\odot$, corresponding to binaries with component mass 2 \msun{} and 1 \msun{}.
This is the boundary beyond which the primary object was allowed to possess higher spin up to $\sim$ 0.98 in 4-OGC, enabling sensitivity to highly spinning sources above this mass.
However, even inside the region with $\mathcal{M}_c$ greater than 1.2 \msun{}, up to $30\%$ of the \ac{SNR} can still be missed in the case of extreme effective spin parameters, with positive spin having a greater loss.
This is because negative effective spin can be compensated by biasing the mass ratio.
For chirp mass less than 1.2 \msun{}, as expected, only the effective spin between $\sim$-0.05 and $\sim$ 0.05 can be recovered with high sensitivity.

The most recent subsolar mass search performed in O3 data by the LIGO-Virgo-KAGRA (LVK) collaboration \citep{2022PhRvL.129f1104A, 2022arXiv221201477T} has overlap in chirp mass and spin magnitude compared with our search, but requires that one component to be subsolar and considers frequencies above 45 Hz. Earlier LVK subsolar searches had considerably less overlap~\citep{PhysRevLett.123.161102,PhysRevLett.121.231103}. Our search covers all existing public data, O1-O3, extends to a higher spin magnitude of 0.95, and we consider frequencies down to 20 Hz. Furthermore, the degeneracy between mass ratio and spin is only partial, so covering a similar chirp mass range does not guarantee sensitivity in our parameter space. 
We find that the LVK template bank will miss $>30\%$ of the detections in our search; this is an optimistic estimation for the loss of detections because signal consistency tests penalize small discrepancies between the signal and template. As a consequence, our search successfully detects the known binary neutron star merger GW170817 and neutron star black hole merger GW200115, as opposed to the LVK subsolar mass searches~\citep{2022PhRvL.129f1104A, 2022arXiv221201477T}.

%\begin{figure}\includegraphics[width=\columnwidth]{lvkff.png}
%\caption{Results of fitting factor from a comparison of simulated waveforms and the template bank from LVK subsolar mass search \citep{2022PhRvL.129f1104A, 2022arXiv221201477T}.}
%\label{fig:lvkff}
%\end{figure}

\section{Search for Low Mass High Spin Binaries}\label{sec:search}

\subsection{search strategy}
\label{sec:method}
The search strategy employed by this work is generally identical to that of the binary neutron star search in 4-OGC \citep{10.3847/1538-4357/aca591}, except for the implementation of a different template bank. 
An open-sourced software \texttt{PyCBC} \citep{pycbc-github} is employed to conduct the search.
We use the \textit{TaylorF2} approximant \citep{0907.0700} to model the gravitational-wave signals and to construct a template bank. \textit{TaylorF2} is a frequency domain waveform model characterizing the inspiral of compact binaries based on a stationary phase approximation.
This approximant is accurate up to the 3.5th post-Newtonian order, taking into account aligned spin effects, and terminates at the Schwarzschild inner stable circular orbit.
This inspiral-only model is appropriate for this search because the merger frequency of our target sources is beyond the sensitive frequency band of LIGO and Virgo.

Since we primarily focus on the mass range consistent with a neutron star, the secondary mass of the bank is chosen to be 1-2 \msun{}.
The primary mass is allowed to extend to 5 \msun{} to also allow for some neutron star black hole binaries.
The component spin is assumed to be aligned with the orbital angular momentum and has an amplitude range from -0.95 to 0.95. 

\begin{figure}
%\begin{center}
\includegraphics[width=\columnwidth]{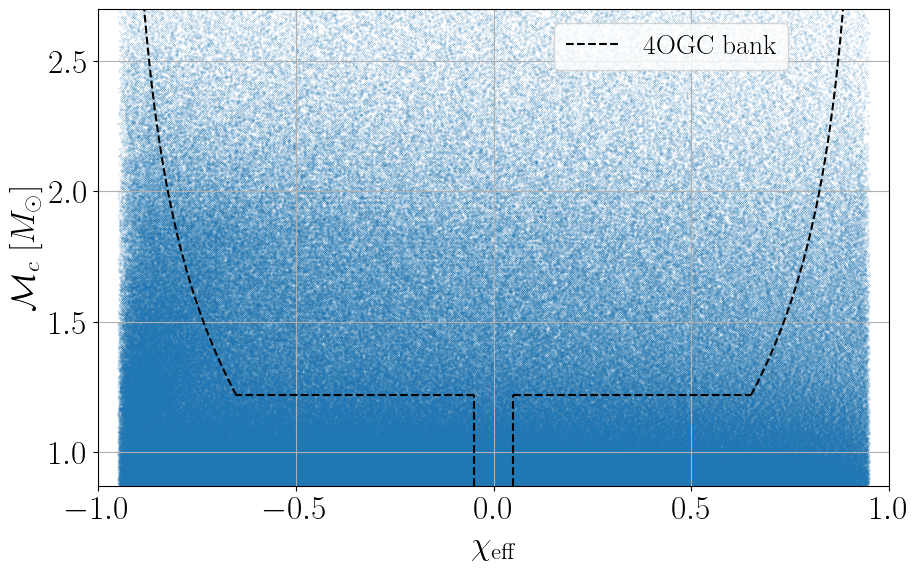}
\caption{The template bank used in this search is plotted in the chirp mass and effective spin plane. 
Each point corresponds to a template, totaling $\sim$700,000.
The dashed line denotes the boundary of the template bank used in 4-OGC as a comparison.
}
\label{fig:bank}
%\end{center}
\end{figure}

The template bank is constructed with a brute-force stochastic placement algorithm \citep{0908.2090}.
In order to enhance the efficiency of generating proposal points iteratively, uniform sampling and sampling from the kernel density estimation using the points already added in the template bank are used alternatively. 
We also pre-record the fitting factor for every newly added template against the existing templates, and use triangular inequality to avoid unnecessary comparison to reduce the computation burden.
Overall, no more than 3\% of \ac{SNR} is lost due to the bank discreteness. 
This bank contains $\sim$700,000 templates, which is plotted in the chirp mass and effective spin space in \cref{fig:bank}.
The dashed line denotes the template bank boundary used in 4-OGC as a comparison.

Matched-filtering is performed over the data from each gravitational-wave detector against all templates. 
Peaks in the resulting \ac{SNR} timeseries are recorded as triggers along with the parameters of the identifying template.
To address non-Gaussian and non-stationary noises, a number of signal consistency tests are conducted, including a $\chi^2$ fit test \citep{10.1103/PhysRevD.71.062001}, a sine-Gaussian $\chi^2$ test of excess power \citep{1709.08974} and test of the \ac{PSD} variability \citep{1908.05644, 10.1088/1361-6382/abac6c}.
Collectively they are combined into a reweighted \ac{SNR} \citep{10.1103/PhysRevD.85.122006, 10.1088/0264-9381/33/21/215004}.

Triggers that are coincident from at least two detectors are combined and assigned a likelihood-inspired ranking statistic \citep{1705.01513,10.1103/PhysRevD.102.022004}.
We do not consider candidates from only a single detector \citep{2019arXiv190108580S, 1910.09528, 2004.10015, 2021CQGra..38i5004A,2203.08545} in this work. The statistic is formulated based on the Neymann-Pearson optimal criterion \citep{0804.1161, 10.1103/PhysRevD.85.122008} to compare the probabilities of astrophysical origin and noise origin.
The likelihood of being classified as noise is determined by empirical analysis by fitting the occurrence rate of triggers' reweighted \ac{SNR} by an exponential function.
The fitting is performed in different parameter regions delineated by the length of the waveform duration, effective spin and mass ratio, which are considered to correspond to noise of different classes. 
The likelihood of being genuine gravitational wave further takes the sensitivity of detectors at the time of triggers into account, as well as the coherence of the \ac{SNR}'s amplitude, phase and time across multiple detectors, compared to an anticipated distribution from a Monte-Carlo simulation of injected signals. 
The determination of statistical significance, also known as the false alarm rate, is accomplished by comparing the ranking statistic with the empirical background distribution, which is obtained by shifting triggers in time between different detectors using an astrophysically forbidden time interval.
Consequently, the statistical properties of background noise can be obtained \citep{10.1088/0264-9381/33/21/215004, 2017PhRvD..96h2002C}.

 \begin{table*}
	\centering
	\caption{The search results of the top 7 candidates. Their inverse \ac{FAR}, component mass $m_{1/2}$, the amplitude of component spin $\chi_{1/2z}$ and the triggered \ac{SNR} in LIGO Hanford, Livingston and Virgo (denoted by H, L, and V) are summarized. }
	\label{table:results}
	\begin{tabular}{lcccccccr} 
		\hline
		Candidates & IFAR(yr$^{-1}$) & $m_1$(\msun{}) & $m_2$ (\msun{}) & $\chi_{1z}$ & $\chi_{2z}$ & SNR(H) & SNR(L) & SNR(V)\\
		\hline
GW170817\_124104 & 17326.85 & 1.66 & 1.15 & 0.09 & -0.09 & 18.39 & 25.98 & - \\
GW200115\_042309 & 3058.95 & 4.88 & 1.88 & -0.40 & -0.17 & 6.33 & 8.90 & - \\
190417\_090325 & 1.01 & 1.12 & 1.08 & 0.33 & -0.84 & 6.34 & 6.90 & - \\
190519\_053312 & 0.29 & 1.77 & 1.25 & 0.72 & 0.63 & 7.38 & - & 5.36 \\
190929\_140800 & 0.29 & 1.01 & 1.00 & 0.01 & -0.38 & 5.25 & 7.06 & - \\
191116\_140156 & 0.18 & 1.24 & 1.12 & 0.72 & 0.73 & - & 6.27 & 6.44 \\
170721\_064738 & 0.14 & 4.22 & 1.89 & 0.63 & 0.47 & 5.39 & 7.06 & - \\
		\hline
	\end{tabular}
\end{table*}

\subsection{Search results}
\begin{figure}
\begin{center}
\includegraphics[width=\columnwidth]{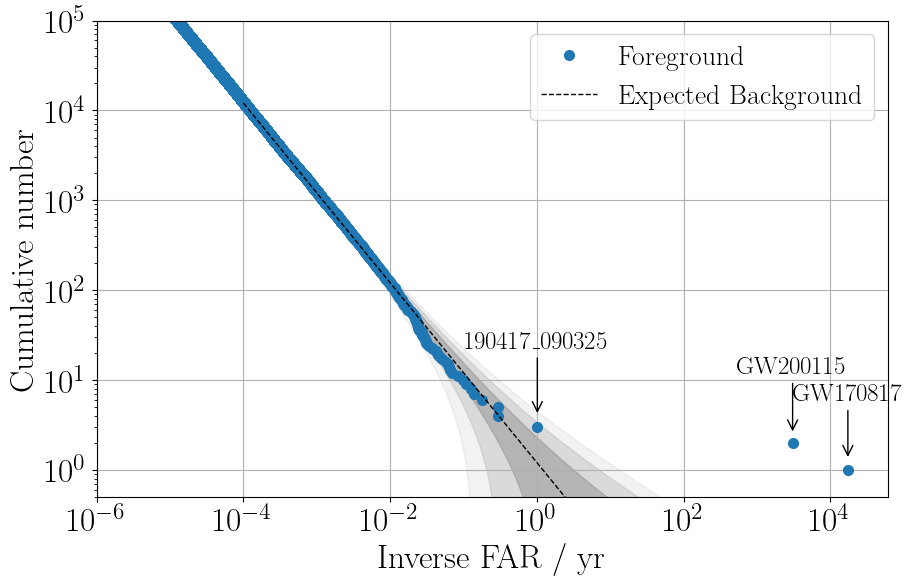}
\caption{The search results are plotted, which illustrates the cumulative number of foreground candidates as a function of inverse \ac{FAR}. 
The dashed line and the shaded region denote the expected distribution of background noise fluctuation and the associated 1, 2 and 3 $\sigma$ from a Poisson process.
We recovered the known events GW170817 and GW200115 with high significance.
The next significant candidate is 190417\_090325.
However, it only has a \ac{FAR} once per 1.01 years thus consistent with a noise fluctuation. 
}
\label{fig:results}
\end{center}
\end{figure}

We conduct a search over the entire publicly available data from the first (O1) to the third (O3) observation runs released by the LIGO and Virgo collaborations, which amounts to $\sim$ 1.2 years of observation time with at least two detectors being online.
\cref{fig:results} illustrates the cumulative number of candidates as a function of the inverse \ac{FAR}.
The parameters from the search for the top candidates are detailed in \cref{table:results}. 

Our search identifies the previously known binary neutron star merger GW170817 and neutron star-black hole event GW200115, with a \ac{FAR} less than once per $\sim10^4$ and $\sim10^3$ years, respectively.
In addition to the known events, the top candidate is 190417\_090325.
This candidate corresponds to a binary with 1.12 \msun{} - 1.08 \msun{}, with component spin 0.33 and -0.84.
However, the \ac{FAR} is only once per 1.01 year.
Based on the amount of observed time, the finding is consistent with noise.
We further checked the data quality and performed parameter estimation for the candidate.
We find that gating a noise transient in LIGO Hanford $\sim$100s before the trigger time reduces the maximum likelihood \ac{SNR} from $\sim15$ to $\sim 8$, suggesting the trigger is likely due to the glitch.
Our investigation did not yield any novel gravitational-wave events.

\section{Constraints on the astrophysical event rate}\label{sec:astro}

In order to translate the null results of the gravitational wave search to an astrophysical event rate upper limit, it is crucial to know the sensitivity of the detector and our search algorithm towards the highly spinning light compact binaries.
The most robust approach is empirically obtaining this knowledge by injecting simulated signals into the detector data stream and replaying the search algorithm.

As a fiducial comparison, we show the results for a 1.4-1.4 \msun{} compact binary in the source frame, a typical mass for binary neutron stars. 
The constraint is not strongly dependent on mass ratio, and the limit can be scaled to other chirp masses as the sensitive distance is $\propto \mathcal{M}_c^{5/6}$ \citep[e.g., see Eq. 2 of][]{2020ApJ...902L..29N}.
To examine the constraints on different spins, we consider six different effective spins, namely $\pm 0.1, \pm 0.5, \pm0.9$.
The sky location, luminosity distance, source orientation, coalescence phase, and polarization angle are chosen to be isotropic or uniformly distributed where applicable.
We generate the simulations, totaling $3\times10^5$ signals, and inject them uniformly into the entire O1 through O3 LIGO and Virgo data. 
The same algorithms as described in \cref{sec:method} are used to identify the simulated sources.

Using the loudest event statistic, which assumes the arrival of signals is a Poisson process, the $90\%$ upper limit for the event rate is given by \citep{0710.0465} 
\begin{equation}
R_{90} = \frac{2.3}{\langle VT \rangle}
\end{equation}
where $\langle VT \rangle$ is the sensitive volume and time of the search at the FAR threshold associated with the loudest candidate, namely 190417\_090325 in this work.
The sensitive volume is computed by a Monte Carlo integral by counting the volume corresponding to the found simulated sources with \ac{FAR} more significant than 190417\_090325 as detected in \cref{sec:search}.

\begin{figure}
\begin{center}
\includegraphics[width=\columnwidth]{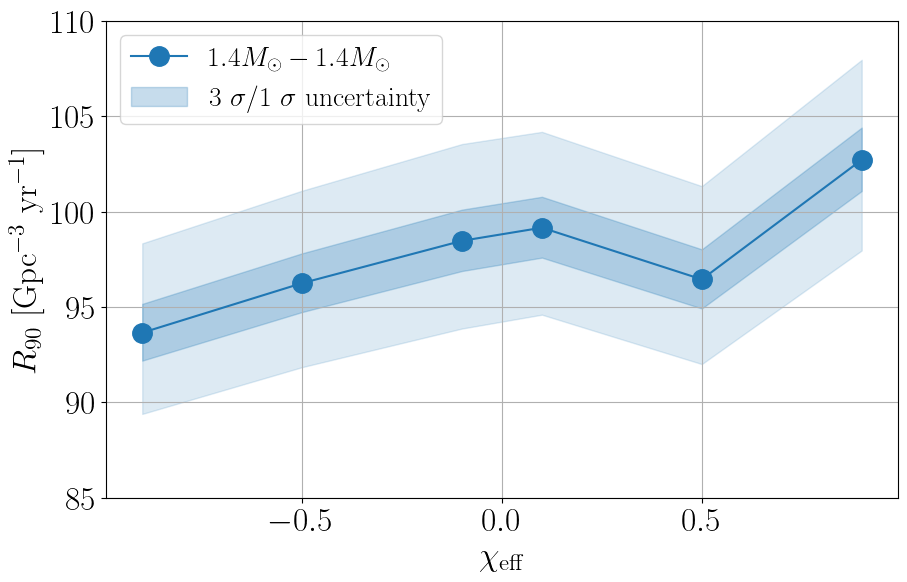}
\caption{The $90\%$ rate upper limit of $1.4-1.4$ \msun{} compact binaries as a function of $\chi_\mathrm{eff}$.
The shaded regions denote the $3\sigma$ and $1\sigma$ Monte Carlo integral uncertainty.}
\label{fig:astro}
\end{center}
\end{figure}

The rate upper limit is depicted in \cref{fig:astro}.
The $90\%$ upper limit of event rate established by this search is $\sim 100 $ Gpc$^{-3}$ year$^{-1}$.
The 3 $\sigma$ and 1 $\sigma$ uncertainty of the Monte Carlo integral is also plotted as the shaded region. This sensitive volume corresponds approximately to an average range of 180 Mpc for each of the six effective spins, respectively. 
We anticipate that the upper limits are not significantly sensitive to the different $\chi_\mathrm{eff}$, with the positive spin being slightly more sensitive due to its longer duration hence a higher \ac{SNR}.
However, we observe that sources with $\chi_\mathrm{eff}=0.9$ are slightly less sensitive than other spins in our search for injections.
A number of factors arising from practice, such as non-Gaussian and non-stationary noises, or the imperfect construction of the template bank, may undermine our anticipation and lead to the observed outcomes.

\section{Discussion and Conclusion}\label{sec:final}

We perform a targeted matched-filtering based search for gravitational waves targeting highly spinning light compact binaries.
In contrast to the parameter space explored in previous searches, which only consider a compact binary with component mass in [1,2] \msun{} and component spin amplitude $[0,0.05]$~\citep{2111.03606,10.3847/1538-4357/aca591}, we extend the range of the amplitude of spin to 0.95.
Potential sources include binary neutron stars resulting from hierarchical mergers~\citep{10.1103/PhysRevD.98.043002} or through unconventional accretion mechanisms~\citep{10.1086/175268}, as well as more exotic objects such as quark stars~\citep{10.1088/0004-637X/728/1/12} or primordial black holes~\citep{2007.06481}. 
It has been found that overly restricting assumptions about binary spin can bias estimation of crucial source parameters~\citep{2111.13619}.
Similarly, searches that do not allow for high spin are less likely to find them.

Our search identifies the previously known gravitational wave events GW170817 and GW200115.
However, no other novel events are detected, suggesting the highly spinning low-mass binaries are relatively rare within the horizon of current detectors. The merger rate is limited to be less than $\sim$ 100 Gpc$^{-3}$ year$^{-1}$ for a fiducial 1.4-1.4 \msun{} compact binary.

The current generation of gravitational-wave detectors is advancing in sensitivity every observing run and are expected to soon achieve their design sensitivity goals.
Meanwhile, there are ongoing projections for the next generation of ground-based detectors, such as Einstein Telescope \citep{2303.15923} and Cosmic Explorer \citep{2306.13745}, with one order of magnitude lower noise compared to Advanced LIGO and better sensitivity at frequencies less than 10 Hz.
As the sensitivity of detectors continues to increase, the prospects for uncovering a hidden population of rare or unusual sources increase in promise. For example, the detection of a sufficiently high-spin binary, could significantly improve our understanding of binary neutron star formation, or support the existence of novel binaries composed of quark stars or primordial black holes.

The method presented in this work can be further optimized by considering finite size effects of the compact components. Tidal interactions are expected to be difficult to measure in the mass range we consider \citep{2105.09151}, however, spin-quadrupole interactions may also have an impact~\citep{1801.09972}. There is also the potential to investigate more exotic sources which require detailed signal modelling and may pose computational or technical challenges if model-dependent parameters require dramatic increases in the number of templates required to conduct a sensitive search.

\section*{Acknowledgements}
YFW and AHN acknowledge the Max Planck Gesellschaft and the Atlas cluster computing team at AEI Hannover for technical support. AHN acknowledges support from NSF grant PHY-2309240.
This research has made use of data, software and/or web tools obtained from the Gravitational Wave Open Science Center (https://www.gw-openscience.org), a service of LIGO Laboratory, the LIGO Scientific Collaboration and the Virgo Collaboration. LIGO is funded by the U.S. National Science Foundation. Virgo is funded by the French Centre National de Recherche Scientifique (CNRS), the Italian Istituto Nazionale della Fisica Nucleare (INFN) and the Dutch Nikhef, with contributions by Polish and Hungarian institutes. 

\section*{Data Availability}

We release the necessary scripts to reproduce this work and the results of the search and upper rate limit in the GitHub repository \url{https://github.com/gwastro/high-spin-light-binary}.

\bibliographystyle{mnras}
\bibliography{reference.bib} % if your bibtex file is called example.bib

\bsp	% typesetting comment
\label{lastpage}
\end{document}